\documentclass{article}

\usepackage{PRIMEarxiv}

\usepackage[utf8]{inputenc} 
\usepackage[T1]{fontenc}    
\usepackage{hyperref}       
\usepackage{url}            
\usepackage{booktabs}       
\usepackage{amsfonts}       
\usepackage{nicefrac}       
\usepackage{microtype}      
\usepackage{lipsum}
\usepackage{graphicx}
\graphicspath{{media/}}     
\usepackage{multirow}
\usepackage{enumitem}
\usepackage{siunitx}

\title{
From 'What-is' to 'What-if' in Human-Factor Analysis: A Post-Occupancy Evaluation Case
}

\author{
  Xia Chen \\
  Georg Nemetschek Institute \\
  Technische Universität München\\
  Munich, Germany\\
  \texttt{x.c.chen@tum.de} \\
  \And
  Ruiji Sun \\
  Center for the Built Environment \\
  University of California, Berkeley \\
  Berkeley, USA\\
  \texttt{ruijis@berkeley.edu} \\
  \And
  Philipp Geyer \\
  Sustainable Building Systems \\
  Leibniz Universität Hannover \\
  Hannover, Germany\\
  \texttt{geyer@iek.uni-hannover.de} \\
  \And
  André Borrmann \\
  Georg Nemetschek Institute \\
  Technische Universität München\\
  Munich, Germany\\
  \texttt{andre.borrmann@tum.de} \\
  \And
  Stefano Schiavon \\
  Center for the Built Environment \\
  University of California, Berkeley \\
  Berkeley, USA\\
  \texttt{schiavon@berkeley.edu} \\
}

\begin{document}
\maketitle

\begin{abstract}

Human-factor analysis typically employs correlation analysis and significance testing to identify relationships between variables. However, these descriptive ('what-is') methods, while effective for identifying associations, are often insufficient for answering causal ('what-if') questions. Their application in such contexts often overlooks confounding and colliding variables, potentially leading to bias and suboptimal or incorrect decisions.

We advocate for explicitly distinguishing descriptive from interventional questions in human-factor analysis, and applying causal inference frameworks specifically to these problems to prevent methodological mismatches. This approach disentangles complex variable relationships and enables counterfactual reasoning. Using post-occupancy evaluation (POE) data from the Center for the Built Environment's (CBE) Occupant Survey as a demonstration case, we show how causal discovery reveals intervention hierarchies and directional relationships that traditional associational analysis misses. The systematic distinction between causally associated and independent variables, combined with intervention prioritization capabilities, offers broad applicability to complex human-centric systems, for example, in building science or ergonomics, where understanding intervention effects is critical for optimization and decision-making.

\end{abstract}

\keywords{Causal Inference \and Counterfactual Reasoning \and Engineering Analysis \and Decision-Making \and Building Science \and Post-Occupancy Evaluation (POE)}

\section{Introduction}

Analyzing numerical data using traditional statistical methods like correlation analysis (Pearson/Spearman) \cite{pearson2011comparison, schober2018correlation} and significance analysis \cite{bentler1980significance, bartlett1950tests} has been the standard approach in many research domains. These techniques effectively identify and quantify associations between variables and assess whether these relationships are statistically significant or not. Furthermore, model interpretation methods like SHAP (SHapley Additive exPlanations) \cite{lundberg2017unified} and LIME (Local Interpretable Model-agnostic Explanations) \cite{ribeiro2016should} provide interpretability for machine learning (ML) models. However, all these methods primarily provide tools to discover patterns (correlations) among variables, allowing researchers to understand the 'what-is' situations within the given data. 'What-is' analysis describes observed correlations in existing data, without considering how changes to variables might affect outcomes under different scenarios.

\begin{figure}[h!]
	\centering
	\includegraphics[width=0.8\textwidth]{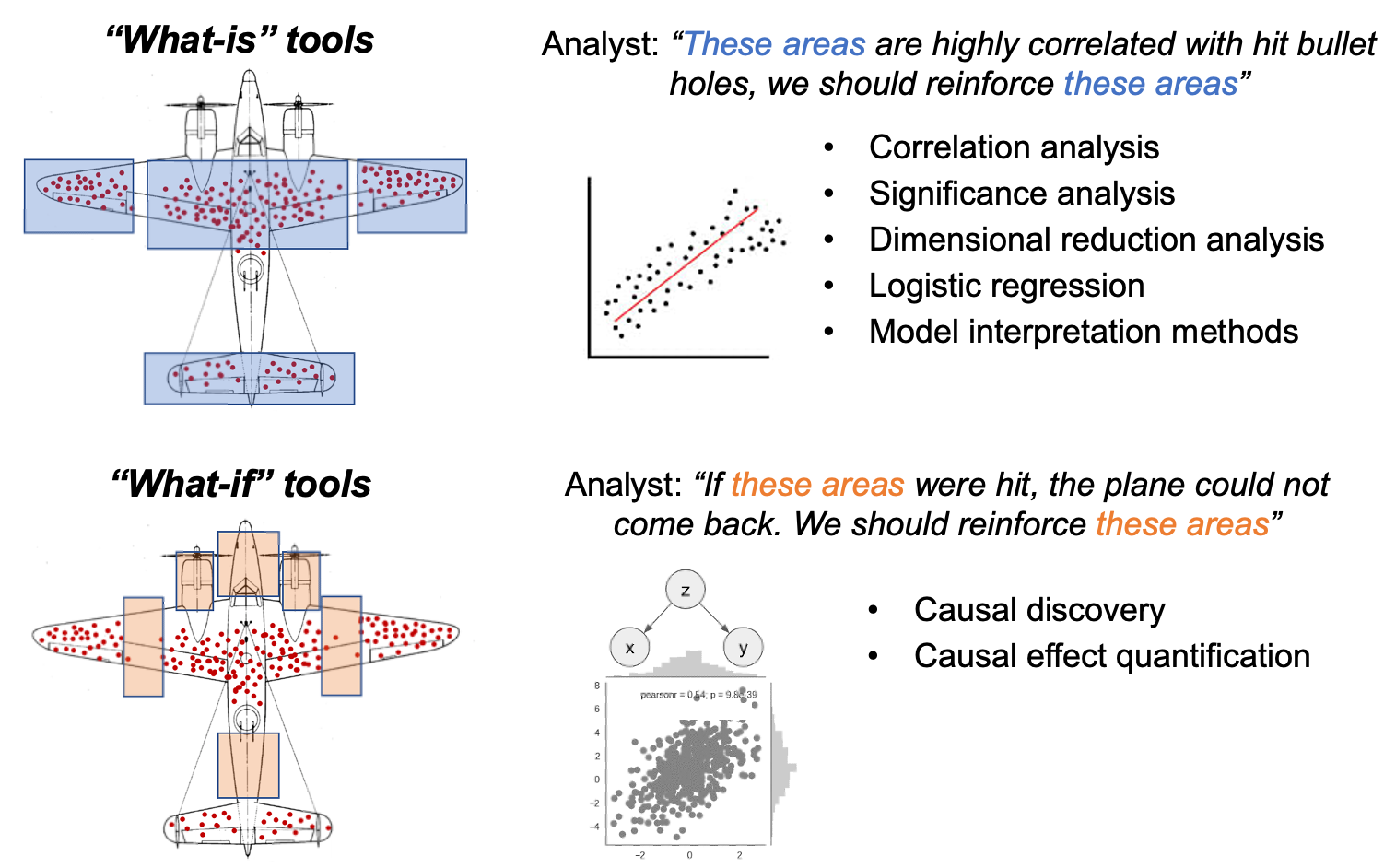}
	\caption{
		Survivorship bias during World War II: Survivor aircraft with skewed bullet hole distribution caused engineers to reinforce less important parts \cite{mangel1984abraham}. The misuse of 'What-is' tools to address 'What-if' questions could lead to spurious conclusions. In contrast, most of the traditional statistical analysis methods are 'What-is' tools  --  \textit{credit of aircraft picture: Martin Grandjean, McGeddon, Cameron Moll, CC BY-SA 4.0}.
	}
	\label{fig:Survivorship_bias}
\end{figure}

However, the core objective of many studies extends beyond understanding 'what-is' (correlational). Instead, researchers aim to understand how modifications to certain variables, like increasing the size of the window, or the type of office, or changing the amount of outdoor air, could modify desired outcomes, thus shifting the focus to 'what-if'. These are causal questions. This distinction reveals a misalignment \cite{lawler2021misalignment} because traditional statistical methods typically assume independence between observations and symmetric associations between variables, without explicitly modeling structural/causal relationships, directional dependencies, thereby failing to account for confounding and colliding factors or the underlying data-generating process that can skew results \cite{christenfeld2004risk, skelly2012assessing}. An intuitive and well-known example is the survivorship bias observed during World War II: Engineers recognized that aircraft returning with wing damage suggested a correlation between bullet holes and wing vulnerabilities. However, the critical insight was that planes with engine damage often did not return at all, meaning that engine vulnerabilities were the true 'what-if' factors to address, as presented in Figure \ref{fig:Survivorship_bias}. The misuse of \textbf{'what-is'} tools to address \textbf{'what-if'} questions could lead to spurious conclusions.

In various domains focused on complex human-centered design and engineering (such as building science or ergonomics), which feature non-linear dynamics that resist straightforward first-principles modeling, a deeper methodological challenge often emerges beyond data availability: the fundamental misclassification of the analytical problem's nature. Analysts frequently employ descriptive, 'what-is' tools to answer causal, 'what-if' questions -- a critical mismatch that can lead to flawed or even counterproductive conclusions. The field of building post-occupancy evaluation (POE) is an example \cite{leaman2006productivity, preiser2015post}: Here, the goal is not merely to statistically describe correlations, but to inform effective interventions to improve occupant satisfaction and well-being. While traditional methods like correlation analysis, significance testing, and regression models have been foundational for assessing and identifying predictors \cite{zhang2022review, sanni2016post, frontczak2012quantitative}, as demonstrated in large-scale surveys \cite{parkinson2023common, xiong2025measuring, graham2021lessons}; other techniques like PCA, cluster analysis are inherently descriptive, even advanced methods like Structural Equation Modeling (SEM)\cite{tekce2020structural, kamaruzzaman2015occupants, loengbudnark2023impact, gul2020investigating}, which test hypothesized causal paths, are primarily confirmatory. Consequently, they risk falling into the same trap as the WWII survivorship bias example: prioritizing the most obvious or frequently reported issues (e.g., noise), while potentially overlooking the true causal drivers of system-wide dissatisfaction (e.g., lighting affecting thermal comfort). This occurs because these methods cannot adequately account for confounding, colliding variables (e.g., building type as a collider \cite{scofield2009leed, newsham2009leed}), or directional relationships. Although emerging research has begun integrating causal inference for 'what-if' scenarios \cite{duhirwe2024causal, sun2025causal}, the broader community has yet to universally adopt this crucial distinction. We aim to demonstrate the value of causal discovery in these contexts through a POE case study, highlighting three key advantages: (1) the simplification of factor analysis by identifying and analyzing causally (non-)associated variables; (2) the ability to prioritize which intervention may be most effective by leveraging the causal variable hierarchy information; and (3) the usage of counterfactual reasoning by providing a causal skeleton to effectively (non-)control factors for unbiased estimation.

In this paper, we demonstrate the value of causal discovery through a POE case study from the Center for the Built Environment (CBE) Occupant Survey database \cite{graham2021lessons}. By applying this methodology, we demonstrate how it can uncover key influencing factors and their hierarchical causal structures, revealing underlying patterns that traditional methods often fail to detect, but which expert opinion can identify. Ultimately, we argue that integrating causal inference methods into existing research and consultancy workflows can provide deeper insights, support more effective decision-making, and help realize the full potential of survey data.

\section{Research background and theoretical basis}
\label{sec:background}

Building POE studies aim to assess the success of building designs and operation, guiding building management, and providing diagnostic feedback for improvements \cite{zimmerman2001post}, which represents a domain where the methodological distinction between descriptive and interventional analysis is critical. As one of the most widely used POE instruments \cite{parkinson2023common, graham2021lessons}, the CBE Occupant Survey collects comprehensive data on occupant demographics, workspace characteristics, and satisfaction across multiple environmental dimensions: thermal comfort, lighting, acoustics, and air quality. However, the resulting datasets, often containing numerous explicit/hidden variables, present significant analytical complexity where traditional correlation-based approaches may mislead intervention priorities. This case study demonstrates how causal discovery can address POE-specific challenges:
\begin{itemize}
	\item \textit{What if certain factor(s) change would significantly improve overall workspace satisfaction?}
\end{itemize}

Table \ref{tab:tab1} describes the differences among common statistical methodologies applied in previous studies \cite{graham2021lessons} and causal inference, along with their addressed questions, limitations, characteristics, and their context in human-factor analysis scenarios.

\begin{table}
\caption{Comparative analysis of statistical and causal inference methodologies: Differentiating 'What-Is' and 'What-If' approaches}
\centering
\begin{tabular}{p{0.14\textwidth}p{0.24\textwidth}p{0.26\textwidth}p{0.26\textwidth}}
\toprule
\textbf{Methodology} & \textbf{Question Addressed} & \textbf{Limitations/Characteristics} & \textbf{Context in Human-factor Analysis} \\
\midrule

\textbf{Correlation Analysis \cite{pearson2011comparison, schober2018correlation, asuero2006correlation}} (Pearson/Spearman) & 
\textbf{'What-Is':} Identifies the strength of the linear relationship or ranks the degree of association between two variables. & 
- Does not consider confounding factors or independence.\newline
- Cannot handle non-monotonic relationships; Pearson correlation is limited to linear associations. & 
Focuses solely on pairwise associations. Identifies correlations between satisfaction factors like lighting and acoustics, but does not involve multi-dynamics \cite{frontczak2012quantitative}. \\

\midrule

\textbf{Significance Analysis \cite{bentler1980significance}} & 
\textbf{'What-Is':} Determines whether the observed relationships between variables are statistically significant. & 
- Requires prior knowledge or predefined hypotheses.\newline  
- Focuses on pairwise analysis, ignoring broader contexts and complex relationships among variables. & 
Identifies the statistical significance of the correlation between overall satisfaction and factors (e.g., temperature, window view, etc.) \cite{kent2024indoor} \\

\midrule

\textbf{Principal Component Analysis \cite{pearson1901liii, abdi2010principal}} (PCA) & 
\textbf{'What-Is':} Reduces dimensionality by identifying key patterns and/or groups of related variables. & 
- The resulting components can be hard to interpret.\newline
- Struggles to establish thematic linkages. & 
Reduces the number of surveyed factors by grouping satisfaction categories \cite{kent2021data}. \\

\midrule

\textbf{Hierarchical Cluster Analysis \cite{bridges1966hierarchical}} (HCA) & 
\textbf{'What-Is':} Identifies clusters of variables with similar characteristics. & 
- Does not reveal directional relationships.\newline
- Biased toward initial groupings. May lead to redundant or overlapping survey questions. & 
Clusters respond by grouping associated characteristics like acoustic privacy, glare/reflections, and air quality \cite{graham2021lessons}. \\

\midrule

\textbf{Logistic Regression \cite{hosmer2013applied}} & 
\textbf{'What-Is':} Identifies factors affecting the likelihood of a binary outcome. & 
- Assumes a linear relationship between predictors and the outcome.\newline
- Cannot capture complex interactions. & 
Explores the relationship between a dependent (e.g., overall satisfaction) and one/more independent factors\cite{frontczak2012quantitative}. \\

\midrule

\textbf{Model Interpretation Methods} (SHAP \cite{lundberg2017unified}, LIME \cite{ribeiro2016should}) & 
\textbf{'What-Is':} Provides model interpretability by explaining predictive model outputs, including non-linear feature contributions and learned associations. & 
- Requires a predefined target variable.\newline
- Accuracy and interpretation depend on the predictive model's performance.\newline 
- Primarily focuses on explaining model decisions rather than identifying relationships among variables. & 
Explains machine learning models' predictions for satisfaction levels based on predefined predictors, but remains focused on model interpretation \cite{tartarini2022personal, zhang2025machine}. \\

\midrule

\textbf{Causal Inference \cite{pearl2009causality, pearl2009causal, rubin1974estimating}} & 
\textbf{'What-If':} Identifies directional relationships to predict the outcomes of potential interventions. & 
- Constructs a causal skeleton representing the directional relationships among variables.\newline
- Accounts for confounding/colliding factors and complex relationships.\newline
- Facilitates counterfactual reasoning and guides decision-making for optimal interventions. & 
Identifies predicted impact based on interventions, providing unbiased counterfactual reasoning (If we do... then...) and guiding decision-making for optimal changes. \\

\midrule

\end{tabular}
\label{tab:tab1}
\end{table}

\subsection{Causal inference: Background}

Causal inference is a framework that seeks to understand cause-and-effect relationships within a dataset by identifying asymmetrical associations between variables \cite{pearl2009causal, chen2022introducing, sun2024causal}. This framework is primarily grounded in Structural Causal Models (SCM), which, for many practical applications, is mathematically compatible with Rubin's potential outcomes framework \cite{pearl2009causal, rubin1974estimating}, though we focus on the SCM approach for its intuitive graphical representation. The methodology is centered on two main stages: causal discovery and causal effect estimation, both of which rely on exploiting specific statistical patterns that distinguish causal from purely associational relationships.

To provide an intuitive understanding of these mechanisms, we briefly explain the core statistical principles, though this necessarily simplifies complex theoretical foundations \cite{pearl2009causal, spirtes2010introduction}: Causal discovery algorithms exploit \textit{conditional independence patterns} that reveal directional relationships \cite{pearl2010causal}. The fundamental insight is that causal relationships create asymmetric statistical signatures: if variable $A$ causes $B$, then conditioning on $A$ should render $B$ independent of $A$'s other causes, but this asymmetric independence pattern typically does not hold when conditioning on $B$ \cite{pearl2009causal, spirtes2010introduction}.

Critical to this process is correctly identifying \textit{confounders} and \textit{colliders} -- two fundamental variable types that can distort causal inference if not properly handled \cite{shrier2008reducing, christenfeld2004risk}. These represent the most basic structural concepts in SCM; for comprehensive treatment of additional causal structures and identification strategies, we refer readers to Pearl's foundational work \cite{pearl2009causal}. \textit{Confounders} are variables that causally affect both $A$ and $B$, creating spurious associations that must be controlled for. \textit{Colliders} are variables caused by both $A$ and $B$, which can induce spurious associations when inappropriately conditioned upon. For instance, consider comparing open-plan versus private offices: open-plan layouts typically feature both uniform overhead lighting (reducing lighting control) and shared HVAC zones (reducing thermal control), while private offices offer both task lighting and individual thermostats. Without controlling for office type, one might incorrectly infer that lighting satisfaction directly causes thermal comfort satisfaction, when both are actually driven by the workspace configuration. Conversely, overall satisfaction might act as a collider if both lighting and acoustics independently influence it, and incorrectly conditioning on overall satisfaction could create artificial associations between lighting and acoustics factors. 

The causal inference methodology proceeds through two complementary stages:

\begin{itemize}
    \item \textbf{Causal Discovery} involves uncovering directional relationships among variables by constructing a causal graph, typically represented as a Directed Acyclic Graph (DAG) \cite{textor2016robust, glymour2019review}. Through systematic conditional independence testing, this process identifies asymmetric associations to detect both direct and indirect causal links \cite{shrier2008reducing}. The resulting hierarchical structure simplifies the exploration space by focusing on the most significant causal associations, enabling researchers to properly identify confounding and colliding variables and understand how different factors interact within the causal network. In our case study, we applied the Greedy Equivalence Search (GES) algorithm \cite{chickering2002optimal, spirtes2010introduction} for this purpose, with implementation details provided in the Appendix.
    
    \item \textbf{Causal Effect Estimation} measures the magnitude of causal impact between variables, enabling predictions of hypothetical interventions \cite{pearl2010causal, rubin1974estimating}. By leveraging the causal structure discovered in the first stage, this process enables unbiased estimation of counterfactual outcomes -- that is, predicting what would happen under different intervention scenarios \cite{hernan2011simpson, neuberg2003causality}. This capability directly supports intervention prioritization by quantifying the potential impact of modifying specific causal factors, thereby providing actionable recommendations.
\end{itemize}

Through this systematic approach, causal inference algorithms construct causal skeletons that map directional dependencies among variables \cite{chickering2002optimal}, providing the structural foundation necessary for moving from descriptive correlation analysis to actionable intervention planning.

\section{Case study}

We selected a typical case from the CBE survey data, a building occupant satisfaction report from a large office building. After filtering out metadata and unrelated features, we refined it to 124 variables (columns) relevant for causal analysis. These variables span major categories including occupant demographics, work background, and detailed satisfaction levels across environmental, ergonomic, and psychological domains (see Table~\ref{tab:data_overview}). Most satisfaction variables are rated on a seven-point Likert scale \cite{likert1932technique}. The complete list of variables, statistical overview, and the processing code are available in Appendix \ref{App:over} and our open-sourced repository\footnote{https://github.com/chenxiachan/Causal-human-factors}.

\begin{table}[htbp]
\caption{Overview of data categories in the CBE satisfaction survey. These variables span major categories including occupant demographics, work background, and detailed satisfaction levels across environmental, ergonomic, and psychological domains (detailed breakdown in Appendix Table \ref{tab:tab2}). Total analytical variables: 124.}
\centering
\begin{tabular}{lp{0.7\textwidth}}
\toprule
\textbf{Category} & \textbf{Description and example variables} \\
\midrule
\textbf{Demographics} & Respondent metadata (e.g., age, gender). \\
\midrule
\textbf{Work background} & Employment context (e.g., time in office, work hours, job description). \\
\midrule
\textbf{Environmental satisfaction} & 7-point Likert scale ratings of: \\
& \hspace{0.5cm}• Air quality (e.g., perception, stuffiness) \\
& \hspace{0.5cm}• Thermal comfort (e.g., temperature, controllability) \\
& \hspace{0.5cm}• Lighting (e.g., electric light, daylight, glare, controllability) \\
& \hspace{0.5cm}• Acoustics (e.g., noise levels, privacy) \\
& \hspace{0.5cm}• ... (and other factors listed in Table S1 of the supplement) \\
\midrule
\textbf{Workspace \& amenities} & Satisfaction with furniture, layout, cleanliness, maintenance, access to amenities, views, food, and water. \\
\midrule
\textbf{Well-being \& policies} & Perceptions of wellness policies, work-life balance, and the impact of the workspace on physical/mental health. \\
\midrule
\textbf{Overall outcomes} & Global satisfaction ratings (e.g., overall workspace satisfaction, intent to stay). \\
\bottomrule
\end{tabular}
\label{tab:data_overview}
\end{table}

\subsection{Simplification of exploration space}

As the first step in our causal analysis, we applied the GES algorithm to identify the causal skeleton from the POE data. This process automatically categorized the 124 factors into two distinct groups: (i) \textit{causally associated} variables, which are connected within the DAG, and (ii) \textit{non-associated} (independent) variables. This causal discovery dramatically reduced the analytical scope: only 28 variables (23\%) were found to be causally associated, while 96 variables (77\%) were classified as independent. This enabled researchers to focus on a substantially smaller, yet more influential, set of factors (see Table \ref{tab:tab3}).

\begin{table}[htbp]
\caption{Causal variable categorization by GES algorithm: Analytical space reduction from 124 to 28 variables (77\% reduction)}
\centering
\begin{tabular}{p{0.48\textwidth}p{0.48\textwidth}}
\toprule
\textbf{Causally Associated Features} & \textbf{Non-Associated Features} \\
\midrule 
\textit{Connected within causal graph; interventions} & \textit{Causally independent; interventions} \\
\textit{likely influence other factors} & \textit{less likely to influence other factors} \\

\midrule

Satisfaction of: & Satisfaction of: \\
 \hspace{0.5cm}• Light & \hspace{0.5cm}• Outdoor working accessibilities \\
 \hspace{0.5cm}• Supportive environment & \hspace{0.5cm}• Amount of individual storage \\
 \hspace{0.5cm}• Furnishment & \hspace{0.5cm}• Wellness policies (vacation, maternity, paternity) \\
 \hspace{0.5cm}• Air quality & \hspace{0.5cm}• Water \\
 \hspace{0.5cm}• Acoustic & \hspace{0.5cm}• Life situation \\
 \hspace{0.5cm}• Layout & \hspace{0.5cm}• Overall office \\
 \hspace{0.5cm}• Thermal & \hspace{0.5cm}• Meeting room \\
 \hspace{0.5cm}• View & \hspace{0.5cm}• Nutrition \\
 \hspace{0.5cm}• Cleanliness & \hspace{0.5cm}• Transportation 
\\
\midrule
Other Attributes: \\
\midrule

 \hspace{0.5cm}• Gender & \hspace{0.5cm}• Personal health inference \\
 \hspace{0.5cm}• Working hours & \hspace{0.5cm}• Floor numbers \\
 \hspace{0.5cm}• Location type & \hspace{0.5cm}• Age \\
 \hspace{0.5cm}• Work type & \hspace{0.5cm}• Location direction 
\\
\midrule
\textit{Focus for causal analysis: 28 variables} & \textit{Independent factors: 96 variables} \\

\bottomrule
\end{tabular}
\label{tab:tab3}
\end{table}

This categorization reveals patterns that align well with building science domain knowledge \cite{kim2012nonlinear}. This convergence between unsupervised algorithmic discovery and expert knowledge provides valuable cross-validation: the algorithm confirms domain theories without prior input, while expert knowledge validates the algorithmic approach's credibility. Most environmental satisfaction variables (lighting, air quality, acoustics, thermal comfort, window views) appear as causally associated, reflecting the interconnected nature of indoor environmental quality. Similarly, key personal attributes (gender, working hours, location type, work type) and workspace characteristics (layout, furniture, supportive environment) emerged as causally connected, aligning with the theoretical understanding of occupant experience drivers.

Conversely, the algorithm identified policy-related variables (wellness policies, personal health status), demographic factors (age, floor number), and peripheral satisfaction measures (water quality, nutrition, transportation) as causally independent. This suggests that while these factors may influence individual satisfaction, they operate largely independently of the core environmental satisfaction network. Notably, '\textit{overall office satisfaction}' appears as non-associated, likely serving as a downstream outcome measure rather than a causal driver, which is often a misconception in survey design.

Critically, this type of independence identification is difficult to achieve through 'what-is' methods. Our validation analysis of POE data (see Appendix \ref{App:tra}) demonstrates this limitation empirically: correlation analysis identified 80 strong associations ($|r|$ > 0.5) and revealed clear variable clusters, but cannot distinguish causal independence from weak associations or provide intervention priorities. The analysis found variables deviating from neutral satisfaction, yet significance testing only ranks problem severity without indicating which improvements would generate cascading effects. Factor analysis groups variables by shared variance but provides no information about causal structure or intervention effects. Only causal discovery can simultaneously identify which variables are structurally connected (and thus require joint analysis) and which are genuinely independent (and can be analyzed separately), while providing the directional information needed for intervention planning.

This data-driven categorization enables a dual analytical strategy: intensive causal analysis on the 28 interconnected variables while treating the 96 independent variables as potential confounders. This reduces computational complexity by 77\% while providing a principled foundation for prioritizing interventions on factors most likely to generate cascading improvements.

\subsection{Prioritization of interventions}

Having identified the 28 causally associated variables from our exploration space reduction, the next critical question becomes: \textit{which interventions should be prioritized?} Traditional statistical methods often fall short by presenting only satisfaction distributions and pairwise correlations, failing to account for how changes propagate through interconnected systems. This leads to the survivorship bias problem introduced earlier -- prioritizing the most obvious or frequently reported issues rather than identifying upstream causal drivers.

Consider our POE context: traditional analysis might prioritize interventions based on dissatisfaction frequency, potentially missing that some low-visibility factors drive multiple downstream problems. For instance, if employees frequently complain about noise but rarely mention lighting issues (e.g., in \cite{graham2021lessons}, Figure 3), a 'what-is' approach would target acoustics first. However, causal analysis might reveal that lighting satisfaction influences multiple satisfaction domains, affecting perceived air quality, workspace comfort, and even acoustic perception through psychological pathways. The critical 'what-if' question becomes: \textit{'Which intervention would generate the most widespread satisfaction improvements?'}

\begin{figure}[h!]
    \centering
    \includegraphics[width=\textwidth]{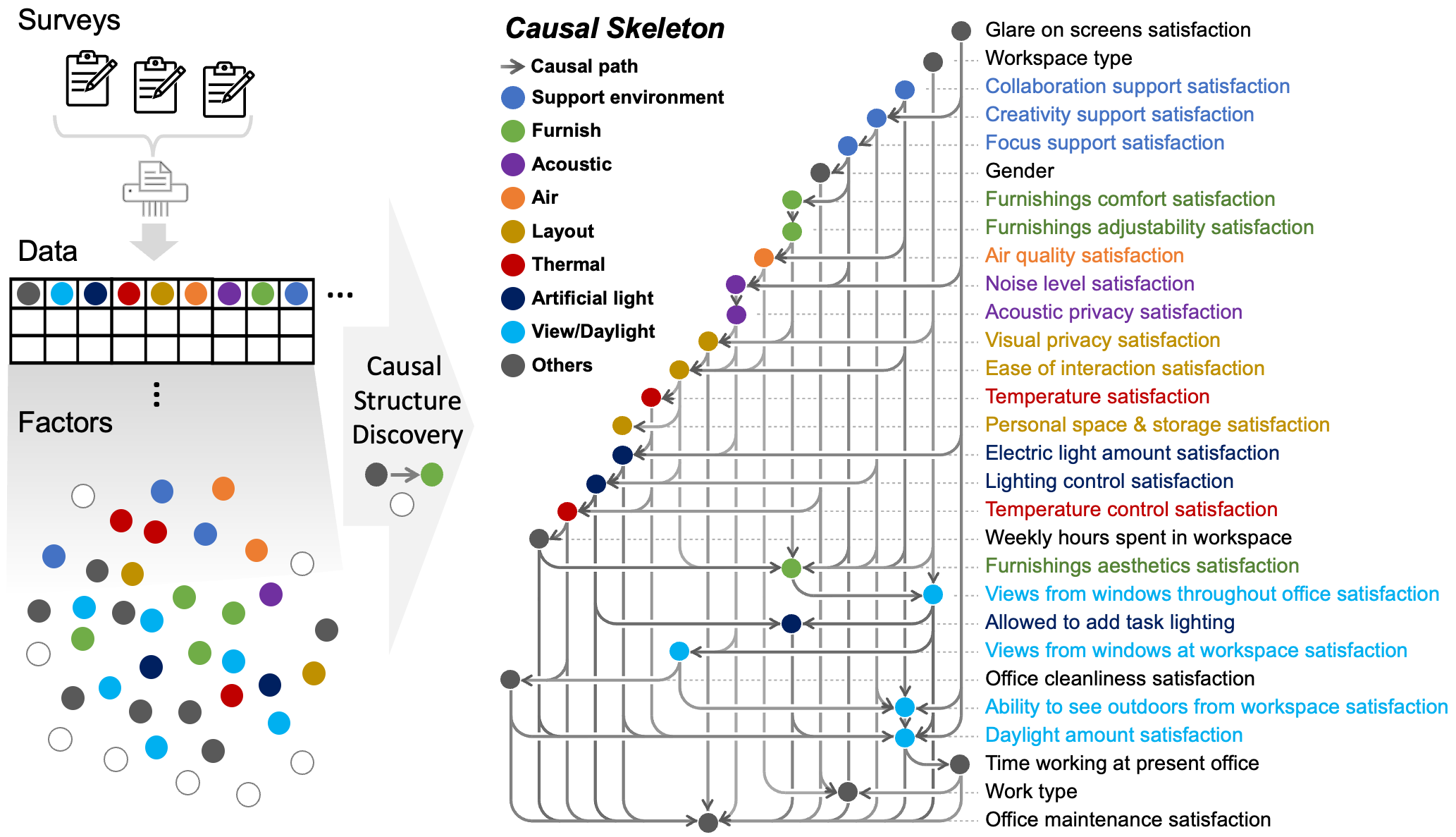}
    \caption{
        From data to a causal skeleton: A visual workflow for uncovering the hidden structure of POE data, discovered by the GES algorithm. As shown on the right, the graph is read from top to bottom, with variables higher in the hierarchy (ancestor nodes) causally influencing those below them (descendant nodes).
    }
    \label{fig:fig2}
\end{figure}

Causal analysis addresses this challenge by establishing hierarchical structures among the 28 causally associated variables (Figure \ref{fig:fig2}), enabling intervention prioritization for effective decision-making. The discovered Causal Skeleton\footnote{Edge orientations are algorithmically determined and should be interpreted as testable hypotheses; implausible directions (e.g., involving fixed demographic variables) likely indicate unmeasured confounders, as discussed in Section \ref{sec:discussion}.} is visualized in Figure \ref{fig:fig2}. For clarity, this figure presents a stylized DAG visualization\footnote{https://github.com/WimYedema/dagviz} of the relationships; the algorithmically generated DAG is included in Appendix Figure \ref{fig:appendix2} for reference.

Within our discovered causal skeleton, we distinguish between \textbf{ancestor} and \textbf{descendant} nodes, representing variables at the top and bottom of the causal hierarchy. Our analysis revealed that \textit{'Satisfaction with glare and reflections on screens'} serves as the most ancestor node, while \textit{'Satisfaction with general maintenance'} appears as the most descendant node.

\subsection{Estimation of causal effects}

\begin{figure}[h!]
    \centering
    \includegraphics[width=\textwidth]{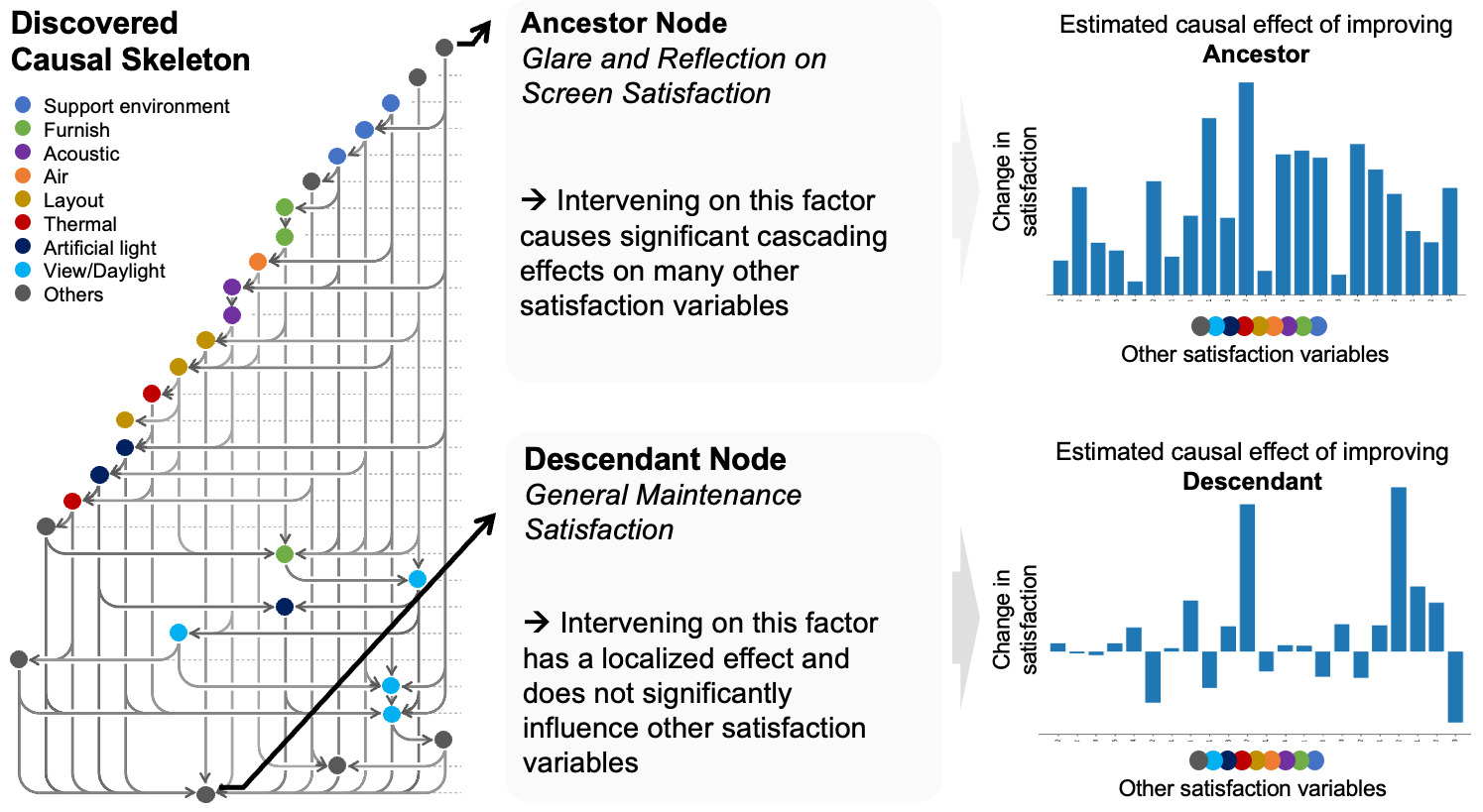}
    \caption{
        Estimating the impact of interventions using causal effects of ancestor and descendant nodes. The left panel shows the causal hierarchy, while the right panel quantifies the average effect of a hypothetical intervention (simulating the do-operator) on other variables by comparing high and low satisfaction groups of the target node. Variable-level results and numerical outputs are provided in the code repository.
    }
    \label{fig:fig3}
\end{figure}

To verify the observations from our causal skeleton, we conducted a numerical analysis using a resampling-based method and analyzed how these changes affected other factors. The core idea of this approach is to estimate the outcome of an intervention, conceptually represented by the \textbf{\textit{do-operator}} in causal theory\cite{pearl2009causal}. For the most ancestral and descendant factors in the skeleton, we targeted them individually and permutationally subsampled them into two groups based on their satisfaction ratings: a 'high satisfaction' group and a 'low satisfaction' group. We then measured the average difference in satisfaction levels across all other variables between these two groups. This process was repeated multiple times to ensure the robustness of the results. As shown in Figure \ref{fig:fig3}, the factors at the top of the hierarchy have a larger causal impact on those at the bottom, and vice versa. This demonstrates that interventions on ancestor nodes yield more widespread effects across the network, confirming their high priority.

This approach revealed that interventions have different effects depending on the variable's position in the causal hierarchy. Improving ancestor factors like glare control tends to generate significant cascading satisfaction improvements across multiple connected factors. Conversely, addressing descendant factors like maintenance shows more localized effects and does not significantly influence the satisfaction of other variables. Understanding these directional relationships enables practitioners to maximize intervention impact by targeting upstream causal drivers rather than downstream symptoms.

\begin{figure}[h!]
    \centering
    \includegraphics[width=\textwidth]{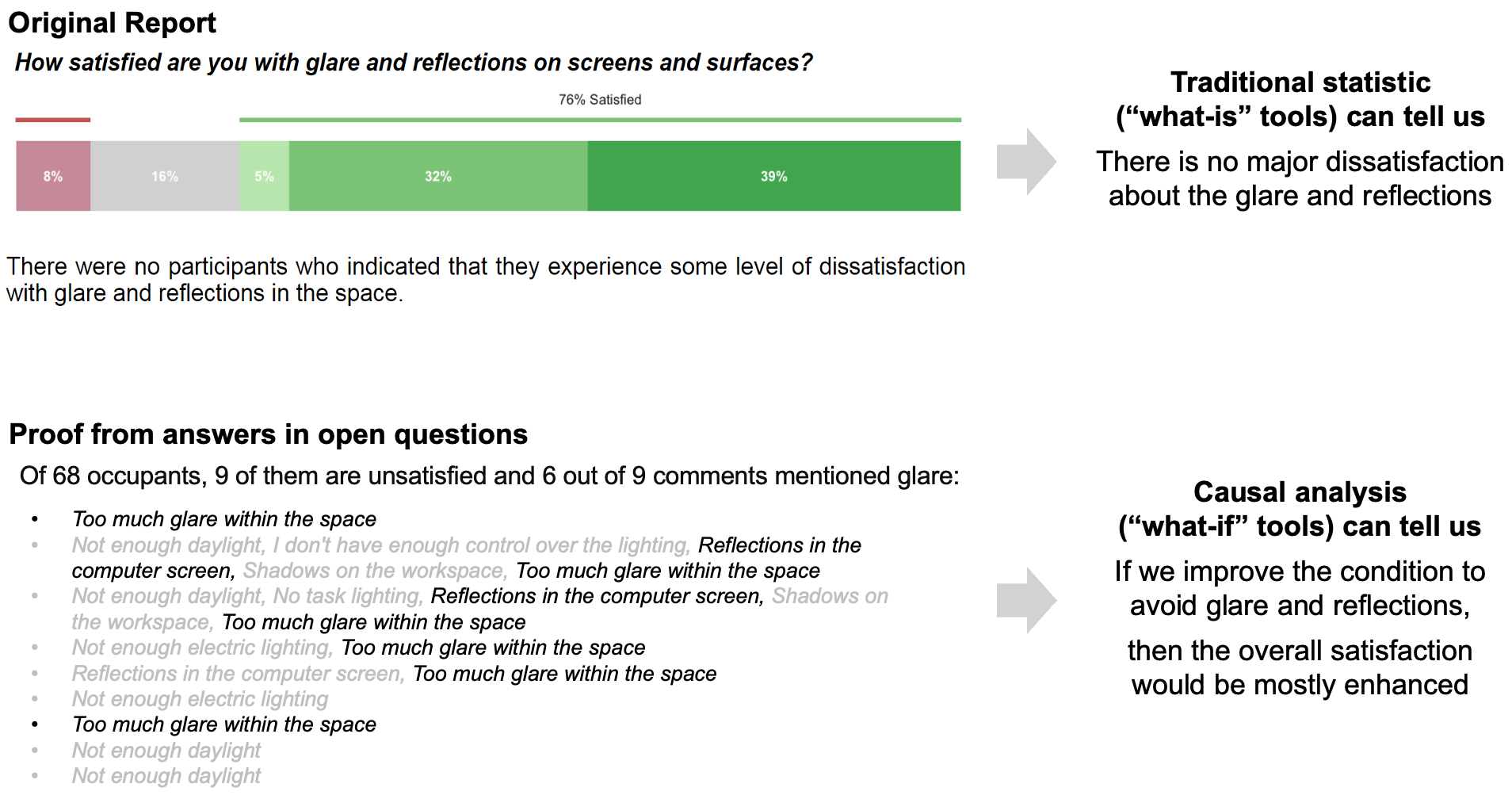}
    \caption{
        Comparison of how the different 'what-is' and 'what-if' tools can lead to oppositely different suggestions for intervention. 
    }
    \label{fig:fig4}
\end{figure}

Moreover, additional validation of the causal skeleton was obtained through the analysis of the open-ended questions in the survey, as presented in Figure \ref{fig:fig4}. During the causal skeleton discovery process from the survey data, such text information was not directly used; thus, there was no information leakage. Notably, we observed that occupants who were most dissatisfied consistently cited glare and reflections as significant issues in their answers to open questions. Although these factors were not the majority source of dissatisfaction in the statistical summary, causal analysis revealed that improving conditions related to glare and reflections could significantly enhance overall satisfaction levels. However, this consequential insight would have been missed if we relied solely on traditional 'what-is' tools like satisfaction distributions. 

Based on our interpretation (domain experts’ perspective), glare and reflections on the screen are an important factor in an office environment. Dissatisfaction with glare and reflections can lead to dissatisfaction in other areas because the occupant's ability to get their job done is impeded. Meanwhile, general maintenance satisfaction is quite broad and can relate to many other elements, but its improvement does not necessarily translate into changes in employees' perception of other factors and may not substantially impact their ability to complete their job.

The causal analysis allowed us to discover the issue with glare as a primary concern that we would have missed with 'what-is' tools, like the summary satisfaction distributions typically reported.

\section{Discussion}
\label{sec:discussion}
The application of causal discovery to POE data revealed directional relationships and hierarchical structures that traditional statistical methods cannot capture. By identifying 'glare and reflections on screens' as a key ancestor node, our analysis uncovered its influential role across multiple satisfaction domains. This insight would likely be overlooked by conventional descriptive approaches, which prioritize high dissatisfaction rates over high-impact causal factors, and would otherwise require costly expert interpretation that cannot be easily automated. The systematic reduction from 124 to 28 causally associated variables demonstrates how causal discovery can focus analytical effort on factors most likely to generate cascading improvements. This distinction between 'what-is' observations and 'what-if' insights represents the core value proposition: moving beyond correlation-based analysis toward intervention-oriented understanding.

While causal discovery offers significant advantages, we acknowledge its remaining challenges inherent to causal discovery from observational data. The presence of hidden or unmeasured variables is a valid critique, as it can complicate the interpretation of results \cite{pearl2010causal, frot2019robust}. The assumption of no unmeasured confounders is particularly challenging in building environments where factors such as outdoor conditions, operational schedules, differentiation on tasks and expectations, and individual psychological and physiological differences could confound observed relationships. These limitations mean our causal structure should be interpreted as testable hypotheses rather than definitive causal claims. It is important to contextualize these challenges; causal inference frameworks have long been instrumental in other complex domains, such as epidemiology and genetics \cite{glymour2019review}, for drawing robust conclusions from observational data. However, as our study shows, even partial knowledge of causal relationships offers valuable guidance for exploration and prioritizing interventions.

Despite these constraints, the discovered structure is well aligned with established the existing literature on occupant satisfaction. This theoretical consistency, combined with validation through open-ended survey responses where dissatisfied occupants consistently cited glare issues, provides partial support for our causal structural assumptions. However, this was just one case study, and therefore, this limits the ability to generalize the results, and the specific finding about glare's prominence may reflect building-specific characteristics rather than universal patterns. Future applications should validate findings across multiple buildings and building types to establish broader applicability. 

We see the effectiveness of causal discovery in revealing intervention hierarchies reinforces as methodological complementarity rather than replacement traditional approaches, at least at this stage. We encourage researchers to first categorize their questions as descriptive ('what-is') or interventional ('what-if') before selecting analytical approaches. Traditional statistical methods remain valuable for descriptive analysis, while causal methods offer unique insights for intervention planning. Expanding this framework to include longitudinal data, controlled interventions, or instrumental variable approaches could strengthen causal identification. The demonstrated ability to generate actionable insights -- even with partial causal knowledge -- suggests significant potential for enhancing POE practice and building performance optimization.

\section{Conclusion}
This study demonstrates the practical value of distinguishing descriptive from interventional questions in human-factor analysis, using post-occupancy evaluation as a case study. Through causal discovery applied to one building assessed with the CBE Occupant Survey, we identified intervention hierarchies that traditional statistical methods would miss, specifically revealing glare and reflections as a high-impact causal factor despite their low visibility in satisfaction complaints. The reduction from 124 to 28 causally relevant variables obtained using a causal discovery algorithm shows how causal methods can focus analytical resources on factors most likely to generate cascading improvements. While our approach has inherent limitations, including cross-sectional data constraints and potential unmeasured confounders, the convergence between algorithmic discovery, domain knowledge, and qualitative validation suggests meaningful practical utility even with partial causal knowledge.

We advocate for a fundamental shift in analytical practice: researchers and practitioners should first categorize their questions as either descriptive ('what-is') or interventional ('what-if') before selecting analytical tools. Traditional statistical methods remain valuable for understanding existing patterns, while causal inference provides unique insights for intervention planning and resource prioritization. The demonstrated ability to reveal actionable intervention strategies from standard survey data, combined with the increasing accessibility of causal discovery algorithms, suggests significant potential for enhancing decision-making across domains where understanding intervention effects is critical for optimization. Causal discovery visualized the underlying causal structures in the survey data and identified critical influencing variables. We think it would be valuable to integrate causal inference methods into research and consultancy workflows, given the low barriers for their adoption.

\section*{Acknowledgments}
This research was partly funded by the industry consortium members of the Center for the Built Environment (CBE), University of California, Berkeley. This work is supported by the author's doctoral research at Leibniz University Hannover, which laid the foundation for this work. The authors thank the Georg Nemetschek Institute on AI for the Built World at the Technical University of Munich for the support provided during the final stages of this work.

The source code and experimental data for this study are publicly available on GitHub at \url{https://github.com/chenxiachan/Causal-human-factors}.

\appendix
\section*{Appendix}
\section{Methodological details of causal discovery}

This appendix provides supplementary technical details on the causal discovery framework employed in this study, intended for readers interested in the underlying methodology and its implementation.

\subsection{Causal graphical models and assumptions}

The foundation of our causal analysis rests on Causal Graphical Models, formally represented as Directed Acyclic Graphs (DAGs) \cite{textor2016robust, textor2015drawing}. In a DAG, nodes represent system variables, and a directed edge from variable $X$ to $Y$ ($X \to Y$) signifies a direct causal influence. The acyclicity constraint enforces the intuitive principle that a variable cannot be its own cause.

This graphical structure encodes a set of conditional independence relationships that can be read from the graph using the \textit{d-separation criterion} \cite{pearl2010causal, neuberg2003causality}. Two sets of nodes $X$ and $Y$ are d-separated (and thus conditionally independent) by a third set $Z$ if every path between a node in $X$ and a node in $Y$ is 'blocked' by $Z$. A path is blocked if it contains:
\begin{itemize}
    \item A chain $A \to M \to B$ or a fork $A \leftarrow M \to B$ where the middle node $M$ is in the conditioning set $Z$.
    \item A collider $A \to M \leftarrow B$ where neither the collider $M$ nor any of its descendants are in the conditioning set $Z$.
\end{itemize}

The validity of inferring a DAG from observational data relies on two key assumptions: (a) the Causal Markov condition, which means that any variable in the DAG is independent of its non-descendants, given its direct parents. (b)  the faithfulness assumption that states that all conditional independence relationships observed in the data are consequences of the d-separation criterion in the true underlying causal DAG. In other words, there are no 'accidental' independencies.

While a violation of faithfulness is theoretically possible, such perfect parameter cancellations are considered highly unlikely in a complex socio-technical system like a building environment, where numerous psychological and physical factors interact. Acknowledging these assumptions reinforces our position that the resulting DAG should be interpreted not as a definitive causal truth, but as a robust, data-driven framework of testable hypotheses. It provides a powerful guide for prioritizing interventions, which can then be validated through targeted, smaller-scale studies.

\subsection{Score-based causal discovery: The GES algorithm}

Greedy Equivalence Search (GES) algorithm is a prominent score-based method for causal discovery \cite{chickering2002optimal}. Unlike constraint-based methods that rely on sequences of individual independence tests, score-based methods search for a DAG (or an equivalence class of DAGs) that best fits the data, as measured by a global scoring function. In our study, we use the \textit{Bayesian Information Criterion (BIC)} (the most common option) as scoring function \cite{neath2012bayesian, mahmood2011structure}. BIC is favored for its ability to balance model fit with model complexity, effectively penalizing overly complex graphs to prevent overfitting. The BIC score is defined as:
$$
\text{BIC} = k \ln(n) - 2 \ln(\hat{L})
$$
where $k$ is the number of parameters in the model, $n$ is the number of samples, and $\hat{L}$ is the maximized value of the model's likelihood function. The goal of the search is to find the graph structure that minimizes the BIC score.

The GES algorithm performs a greedy search over the space of causal graph equivalence classes in two distinct phases. The first one is the Forward Equivalence Search (FES). Starting with an empty graph (no edges), the algorithm iteratively adds the single edge that results in the largest improvement (decrease) in the BIC score. This phase continues until no further single-edge addition can improve the score. The second is the Backward Equivalence Search (BES). Starting with the graph obtained from the FES phase, the algorithm iteratively removes the single edge that results in the largest improvement in the BIC score. This phase continues until no further single-edge removal can improve the score, helping to prune any spurious edges added during the forward phase.

The final output is a representative DAG from the equivalence class that is locally optimal with respect to the BIC score, given the greedy search heuristic. While GES cannot guarantee finding the globally optimal structure, it is a highly effective and widely used algorithm for recovering causal relationships from observational data.

\section{Empirical statistical analysis of POE data}
\subsection{Data overview}
\label{App:over}

\begin{figure}[htbp]
    \centering
    \includegraphics[width=\textwidth]{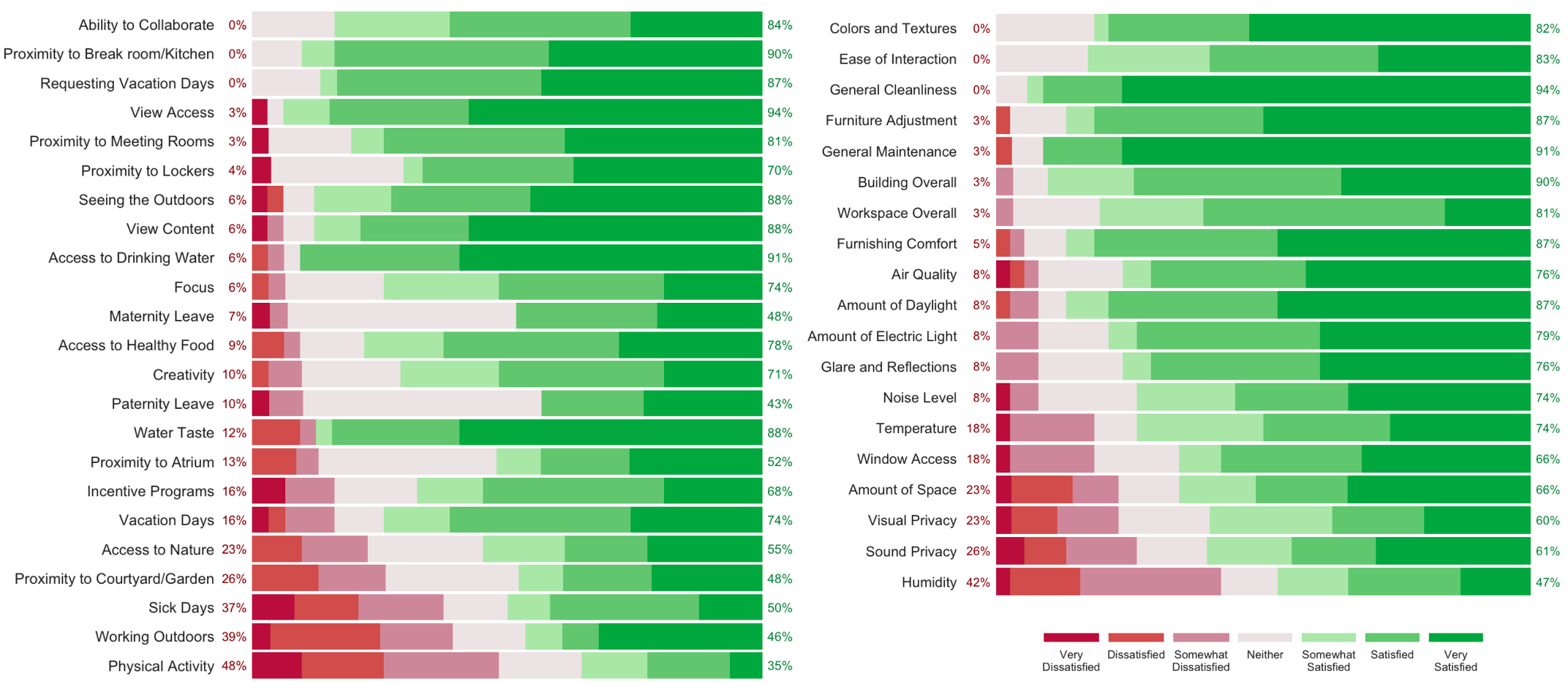}
    \caption{
        Statistical summary of the satisfaction votes in the case study.
    }
    \label{fig:appendix1}
\end{figure}

Building on the survey data overview in Table \ref{tab:tab2} and the statistical summary of satisfaction votes in Figure \ref{fig:appendix1}, our empirical analysis of the validation dataset reveals both the capabilities and critical limitations of traditional 'what-is' statistical methods in POE contexts.

\begin{table}[htbp]
\caption{Overview of CBE satisfaction survey data: a large office building (filtered from 130 to 124 analytical variables)}
\centering
\begin{tabular}{p{0.30\textwidth}p{0.55\textwidth}p{0.10\textwidth}}
\toprule
\textbf{Category} & \textbf{Key variable types} & \textbf{Count} \\
\midrule
\textbf{Demographics \& work} & $\bullet$ Age $\bullet$ Gender $\bullet$ Time at office $\bullet$ Hours per week $\bullet$ Work description & 7 \\
\midrule
\multicolumn{3}{l}{\textbf{Environmental satisfaction (7-point Likert scale):}} \\
\midrule
\textbf{Air quality} & $\bullet$ Satisfaction ratings $\bullet$ Dissatisfaction problem levels $\bullet$ Contributing factors & 7 \\
\textbf{Thermal comfort} & $\bullet$ Temperature satisfaction $\bullet$ Control ability $\bullet$ Seasonal preferences $\bullet$ Problems & 7 \\
\textbf{Lighting} & $\bullet$ Electric light $\bullet$ Daylight $\bullet$ Glare/reflections $\bullet$ Control $\bullet$ Task lighting & 8 \\
\textbf{Acoustic} & $\bullet$ Noise levels $\bullet$ Communication privacy $\bullet$ Problem identification & 5 \\
\textbf{Furniture} & $\bullet$ Comfort $\bullet$ Adjustability $\bullet$ Colors/textures $\bullet$ Specific problems & 7 \\
\textbf{Workspace layout} & $\bullet$ Space amount $\bullet$ Visual privacy $\bullet$ Interaction ease $\bullet$ Problem details & 10 \\
\textbf{Cleanliness \& maintenance} & $\bullet$ General cleanliness $\bullet$ Maintenance quality $\bullet$ Problem identification & 6 \\
\midrule
\textbf{Location \& amenities} & $\bullet$ Floor/area $\bullet$ Window proximity $\bullet$ Views $\bullet$ Nature access $\bullet$ Amenity proximity & 16 \\
\textbf{Water \& food} & $\bullet$ Water taste/access $\bullet$ Healthy food options $\bullet$ Problem factors & 8 \\
\textbf{Wellness policies} & $\bullet$ Vacation/sick days $\bullet$ Leave policies $\bullet$ Health incentives $\bullet$ Usage tracking & 9 \\
\textbf{Physical \& mental health} & $\bullet$ Health ratings $\bullet$ Work environment impact $\bullet$ Stress levels & 9 \\
\textbf{Physical Activity} & $\bullet$ Time allocation (sitting/standing/walking) $\bullet$ Activity satisfaction & 7 \\
\textbf{Support \& Life satisfaction} & $\bullet$ Workspace support (focus/creativity/collaboration) $\bullet$ Life satisfaction $\bullet$ Job satisfaction & 14 \\
\textbf{General workspace} & $\bullet$ Overall workspace satisfaction $\bullet$ Building satisfaction $\bullet$ Work enhancement & 4 \\
\midrule
\multicolumn{2}{l}{\textbf{Variable used}} & \textbf{124} \\
\multicolumn{3}{l}{\footnotesize \textit{Note: Filtered from 130 total columns by removing 5 administrative fields}} \\
\multicolumn{3}{l}{\footnotesize \textit{and retaining all remaining survey variables.}} \\
\bottomrule
\end{tabular}
\label{tab:tab2}
\end{table}

\begin{figure}[htbp]
    \centering
    \includegraphics[width=\textwidth]{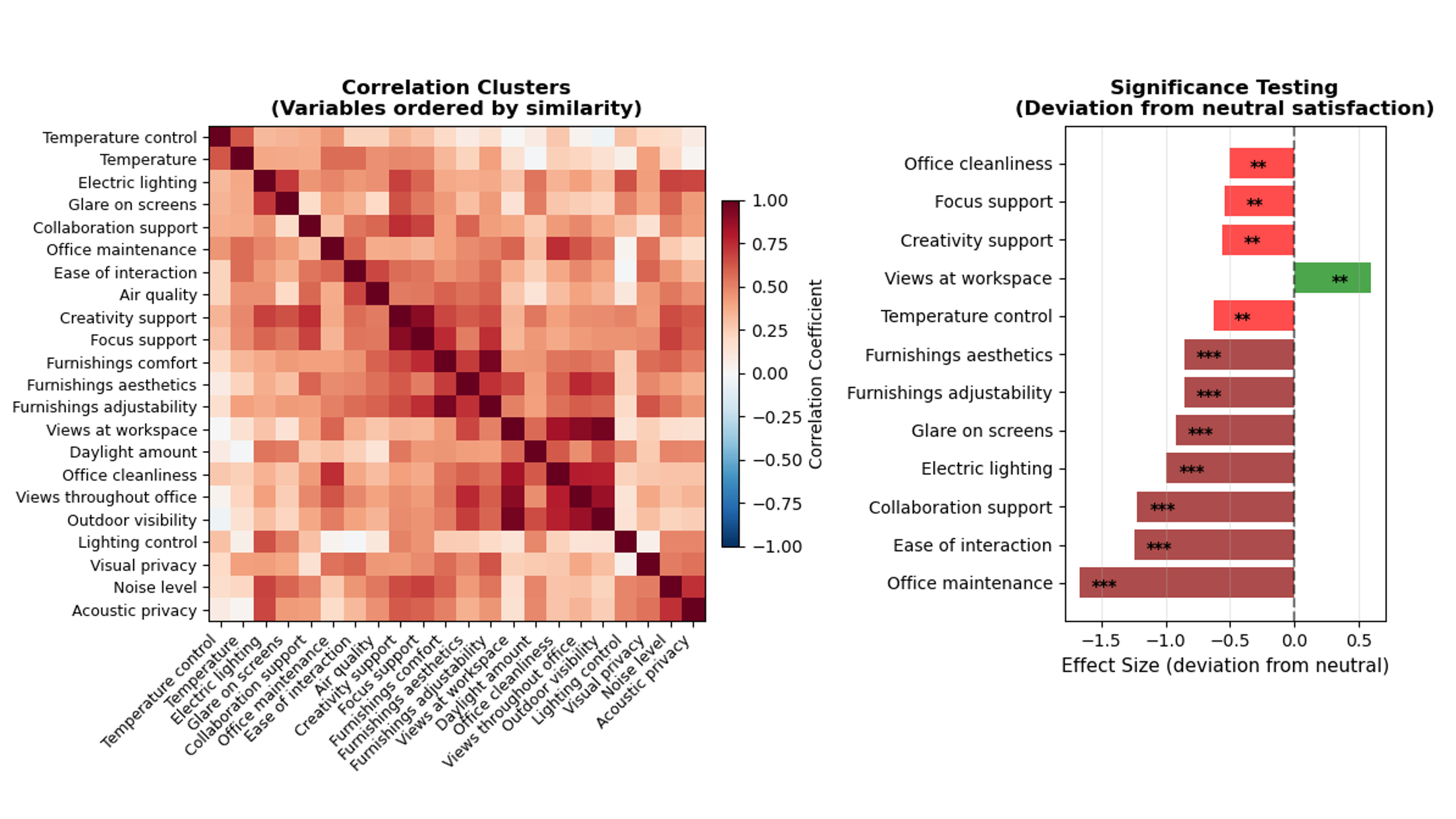}
    \caption{
        Traditional statistical methods applied to POE validation data. Left panel shows correlation analysis revealing variable clusters ordered by similarity, with correlation coefficients ranging from -1 to +1. Right panel shows significance testing results measuring deviation from neutral satisfaction, with effect sizes and statistical significance levels indicated by asterisks (*p<0.05, **p<0.01, ***p<0.001).
    }
    \label{fig:appendix2}
\end{figure}

\subsection{Analysis results}
\label{App:tra}
Our empirical analysis of the validation dataset reveals both the capabilities and critical limitations of traditional 'what-is' statistical methods in POE contexts. The correlation analysis (Figure \ref{fig:appendix2}, left panel) successfully identified 80 strong correlations ($|r|$ > 0.5) among the satisfaction variables, with the strongest associations occurring between conceptually related factors: views at workspace and outdoor visibility (r = 0.954), furnishings comfort and adjustability ($r$ = 0.952), and workspace support factors ($r$= 0.906). The hierarchical ordering reveals clear clustering patterns where related variables group together, confirming that traditional methods excel at identifying associational structures.

Significance testing (Figure \ref{fig:appendix2}, right panel) measured how each variable's mean satisfaction deviates from the neutral point (4 on the 7-point Likert scale). A negative effect size indicates dissatisfaction (mean below neutral), while a positive effect size indicates satisfaction (mean above neutral). Office maintenance showed the largest negative effect size (-1.67, $p$ < 0.001), indicating substantial dissatisfaction. In contrast, views at workspace showed a positive effect size, reflecting that occupants were generally satisfied with this aspect. Traditional methods effectively quantify the severity of dissatisfaction and rank problems by statistical significance.

However, these results simultaneously demonstrate the fundamental limitations that motivate our causal approach. The correlation clusters, while statistically robust, provide no directional information to guide intervention planning. The strongest correlation (0.954) between views and outdoor visibility merely confirms an obvious relationship without indicating causation or intervention potential. More critically, the significance testing identifies office maintenance as the most problematic area, but this finding could represent a downstream symptom of upstream causal factors rather than a root cause requiring intervention.

Finally, we present the GES-generated DAG based on CBE POE data, as shown in Figure \ref{fig:appendix2}.

\begin{figure}[htbp]
    \centering
    \includegraphics[width=\textwidth]{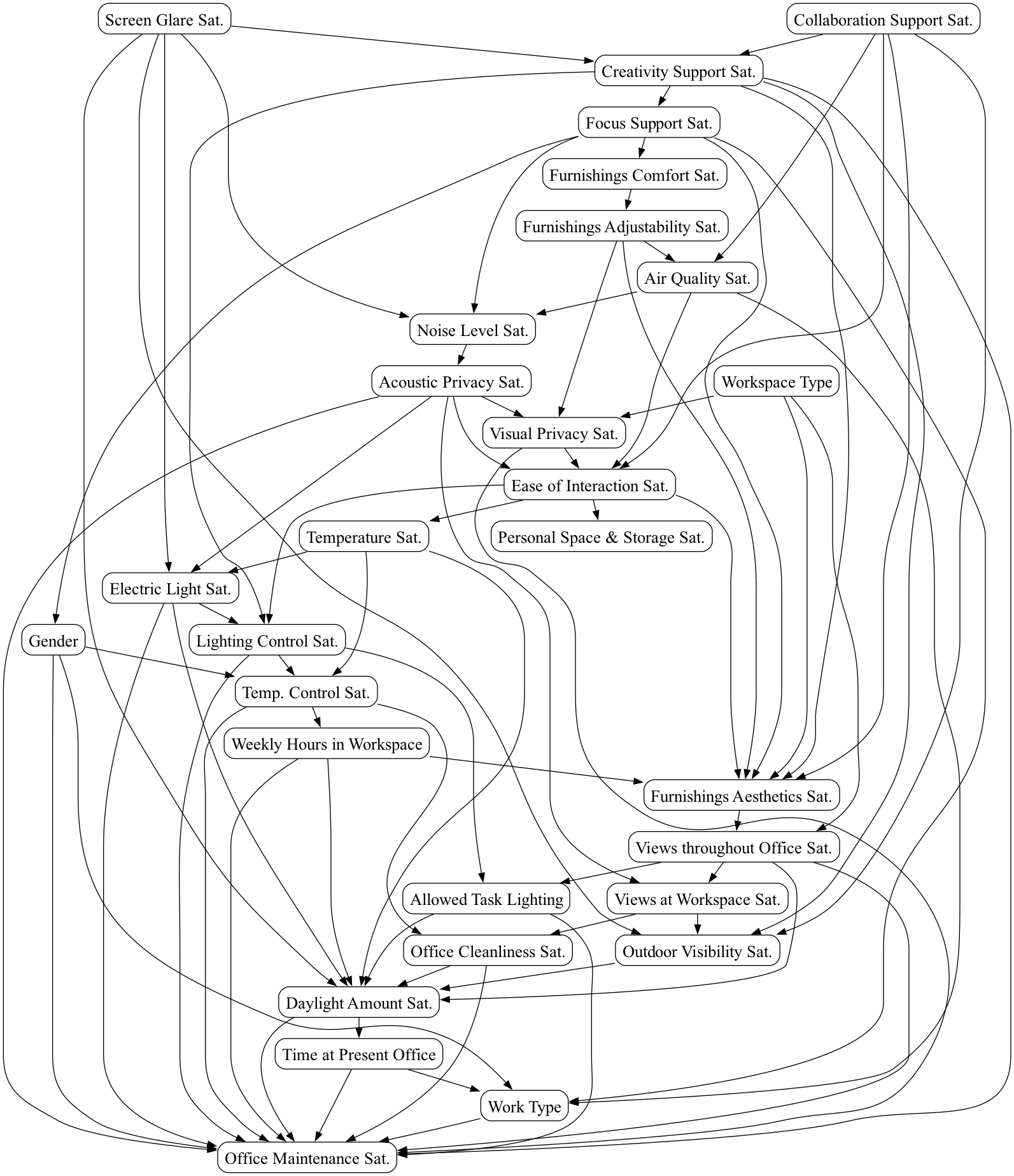}
    \caption{
        Algorithmically-generated DAG from CBE POE data.
    }
    \label{fig:appendix2}
\end{figure}

\clearpage

\bibliographystyle{unsrt}  
\bibliography{references}

\end{document}